# A meteorite crater on Mt. Ararat?

V.G.GURZADYAN  (1)  and  S.AARSETH  (2)

1. Yerevan Physics Institute and Yerevan State University, Yerevan, Armenia; gurzadya@icra.it

2. Institute of Astronomy, Cambridge, UK;  sverre@ast.cam.ac.uk

We briefly report on a crater on the western slope of Mt.Ararat . It is located in an area closed to foreigners at an altitude around 2100m with geographic coordinates  39˚ 47' 30''N,  44˚ 14' 40''E. The diameter of the crater is around 60-70m, the depth is up to 15m. The origin of the crater, either of  meteorite impact or volcanic, including the evaluation of its age, will need detailed studies .

In August  2004 while climbing Mt. Ararat we came across a well outlined crater on its western slope. Ararat is best known from the Bible as the mountain harboring the Noah's Ark, and has attracted many expeditions to search for the Ark including e.g. with participation of Lunar astronaut James Irwin.

Ararat is an isolated volcano of 5156m absolute (above sea level) and of 4300m relative altitude from the valley bottom, of 130 km circumference, with over 30 glaciers. The first recorded summit ascent was by the 1829 expedition of Parrot of Dorpat University (now Tartu, Estonia) [1].

We had only a brief opportunity for the crater, as our inspection and taking of a photo was disturbed by our local guides; the western and northern slopes of Ararat are closed to foreigners[1]. As a result an entire area remains practically unexplored; for example, we had a chance to locate medieval ruins, which later appeared to be [2] a V-VI century Armenian basilica unknown to experts and not entered into the catalogs.

The crater has well outlined borders, more rocky on the western side, as seen in the photo. Although the volcanic origin of the crater cannot be excluded, it is definitely unique for the structure of the slopes of Ararat, and at least no report for such a phenomenon is found both for Ararat and for thoroughly surveyed nearby volcanoes and volcanic ranges (Aragatz, Geghama, etc).

---

[1] Later Turkish military authorities granted us an exceptional permission to climb the mountain from the North.

In sum, what one can stress absolutely surely:

1. the clearly outlined round shape of the crater;

2. clear cut of the borders;

3. it cannot be a result of a glacier erosion; as mountaineers, we recognize well structures formed due to glaciers, also the surrounding soil was not of a *morene* type. Moreover, 2100m is far below the glacier line (now at about 3800m), even taking into account possible climate variations.

Whether the crater is of meteorite impact or of volcanic origin, can be clarified after detailed inspection, and attracting attention to that is the aim of this Note.

1. F.Parrot, *Reise zum Ararat*, Berlin, 1834.
2. V.G.Gurzadyan, V.Harutyunian, *Historical-Philololigal Journal of Armenian Academy of Sciences,* **172,** 206, 2006.

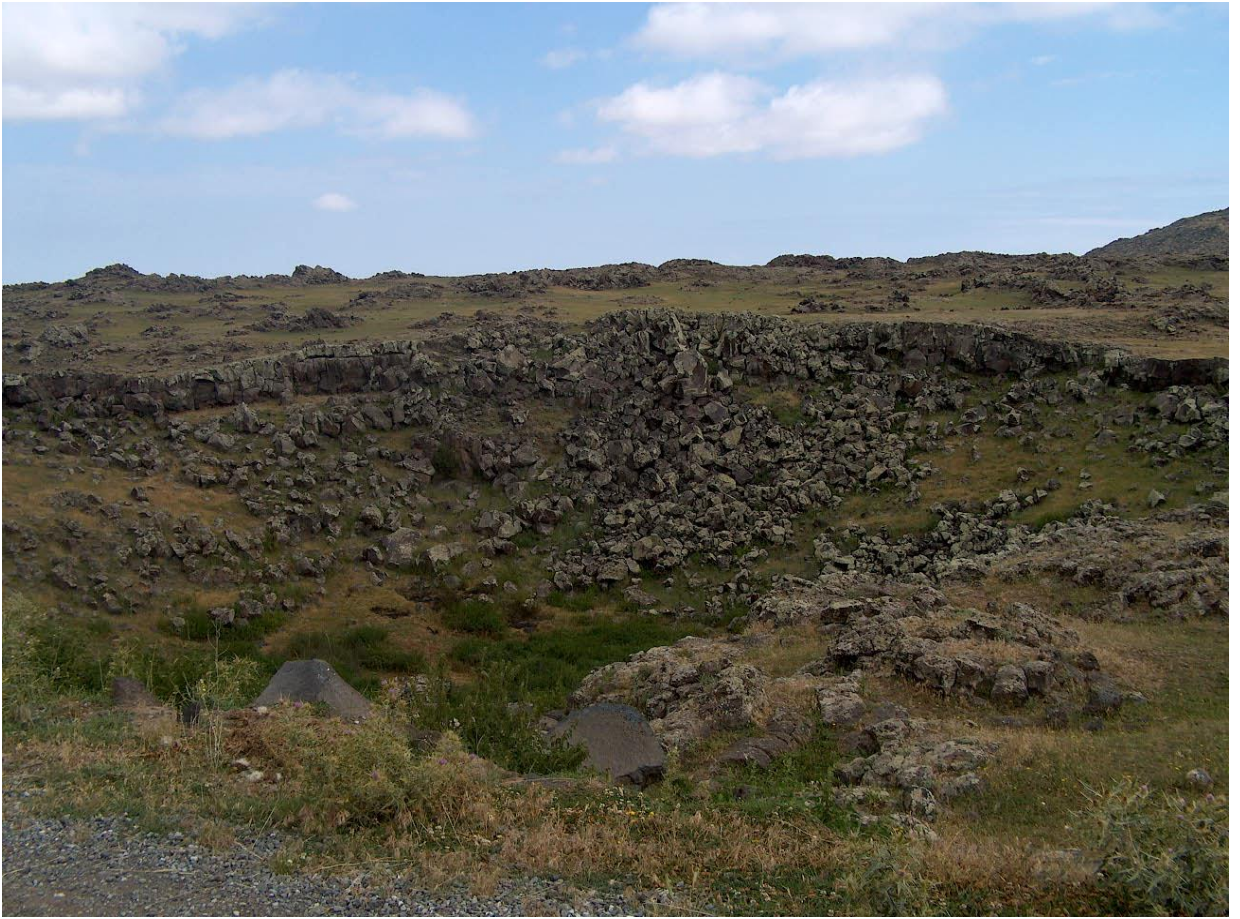

The view of the crater from the East (photo by V.Gurzadyan).